\documentclass[12pt,a4paper]{article}

\usepackage{amssymb}
\usepackage{amsmath, epsfig, natbib, graphics, subfigure}

\newcommand{\la}{\label}
\newcommand{\be}{\begin{equation}}
\newcommand{\en}{\end{equation}}

\renewcommand{\vec}[1]{\boldsymbol{#1}}
\newcommand{\ii}{\textrm{i}}
\newcommand{\ee}{\textrm{e}}

\def\halft{{\textstyle\frac{1}{2}}}

\begin{document}

\title{Bimodular rubber buckles early in bending}

\author{Michel Destrade$^a$, Michael D. Gilchrist$^a$, \\Julie A. Motherway$^a$, Jerry G. Murphy$^b$\\[12pt]
$^a$School of Electrical, Electronic, and Mechanical Engineering, \\
University College Dublin,\\
 Belfield, Dublin 4, Ireland\\[12pt]
$^b$Department of Mechanical Engineering, \\
Dublin City University,\\ Glasnevin, Dublin 9, Ireland.}


\date{}
\maketitle

\numberwithin{equation}{section}


\bigskip

\begin{abstract}

A block of rubber eventually buckles under severe flexure, and several axial wrinkles appear on the inner curved face of the bent block. 
Experimental measurements reveal that the buckling occurs earlier ---at lower compressive strains--- than expected from theoretical predictions.
This paper shows that if rubber is modeled as being bimodular, and specifically, as being stiffer in compression than in tension, then flexure bifurcation happens indeed at lower levels of compressive strain than predicted by previous investigations (these included taking into account finite size effects, compressibility effects, and strain-stiffening effects.)
Here the effect of bimodularity is investigated within the theory of incremental buckling, and bifurcation equations, numerical methods, dispersion curves, and field variations are presented and discussed. 
It is also seen that Finite Element Analysis software  seems to be unable to encompass in a realistic manner the phenomenon of bending instability for rubber blocks.

\end{abstract}

\bigskip


\newpage


\section{Introduction}


Rubber is a highly deformable solid. When we are presented with a rubber band or a rubber string, we almost automatically subject it to a large \emph{stretch} to test its extensibility and then sometimes consider its behavior in \emph{torsion}. However, a \emph{thick, stubby, rubber block} does not lend itself naturally to these two deformations, nor to compression or shear. In fact, one can say that large \emph{bending} (or \emph{flexure}) is the favorite mode of deformation of rubber blocks. Moreover, many engineering and industrial devices rely on the excellent bending characteristics of rubber blocks and subject them to countless bending/straightening cycles. The importance of bending of blocks in applications is the motivation for our study of their stability.

Gent and Cho (1999) subjected blocks of natural rubber to severe bending until they saw creases appearing on the inner bent face. We might expect these axial wrinkles to form on this bent face because of our intuitive notion of a region of tension and a region of compression in the neighborhood of the outer and inner faces of a bent block, respectively. 
Once circumferential line elements contract up to a certain \emph{critical stretch} $\lambda_\text{cr}$, say, the experiments of Gent and Cho thus suggest that incremental wavy static deformations can be superposed on the primary, non-linear  bending deformation, signaling the onset of instability (in the linearized sense). Gent and Cho (1999) assumed that the inner face bending instability should occur at the same critical stretch as that of surface instability of an incompressible half-space. For the neo-Hookean material, they therefore expected the inner face to buckle at $\lambda_\text{cr} = 0.54$, and were surprised to see it occur earlier, at $\lambda_\text{cr} = 0.65 \pm 0.07$.

Their prediction relied on several assumptions. Several studies have  tested these assumptions to discover whether modifying any of them would significantly increase the value of $\lambda_\text{cr}$. We now look at these assumptions in turn. First, the incompressibility assumption: does the introduction of slight compressibility (as is suitable for natural rubbers) change the critical stretch of surface instability? The answer, recently established by Murphy and Destrade (2009), is that slight compressibility has little quantitative effect on the value of the critical ratio of compression for surface instability; if anything, it makes the half-space more stable. Next, the half-space assumption: does the introduction of a finite size for the block make a big difference? The answer is also no, as shown by Haughton (1999) and Coman and Destrade (2008), with the critical stretch slightly increased from 0.54 (half-space in plane strain compression) to 0.56 (block in large bending), with very little variation with the block's dimensions. The final assumption is that the neo-Hookean strain-energy function is an adequate model of natural rubber. In fact, the Mooney-Rivlin material, and also the most general third-order elasticity model of incompressible materials, are all equivalent to the neo-Hookean material in the case of plane strain bending (Goriely et al., 2008; Destrade et al., submitted) so that quite a wide variety of materials reduce to the neo-Hookean form, although it is characterized by only one material parameter, the initial shear modulus. Moreover, Gent and Cho (1999) are correct in their assertion that it might not be `necessary to consider stress-strain relations incorporating finite-extensibility effects' for this problem. Destrade et al. (2009a) studied the impact of the strain-stiffening effect on bending instability. Their conclusion is that it does promote instability for materials which stiffen early (typically at 10-20\% extension stretch) such as biological soft tissues, but that it does not affect the bending instability of materials which stiffen at large stretches (at 200-600\%, say) such as natural rubbers.

In this paper, we investigate the influence of a completely different assumption, one that is usually implicit in the study of non-linearly elastic materials and, in particular, implicit in the work of Gent and Cho (1999). This assumption is that the mechanical behavior of rubber is the same in tension as in compression. 

Intuitively, we expect that there are regions of tension and regions of compression in a deformed solid; however it is a far from trivial task to provide a rigorous definition of these terms, especially in non-linear elasticity, as can be seen in the elegant study by Curnier et al. (1999).
Here we focus on the \emph{flexure} of a block, where intuition suggests that the region near the outer face of the bent block is a region of ``tension'', and the region near the inner face is a region of ``compression''.
In these regions, line elements originally aligned along the length of the bar are extended and contracted, respectively.
 
\begin{figure}
\begin{center}
\includegraphics*[width=12cm]{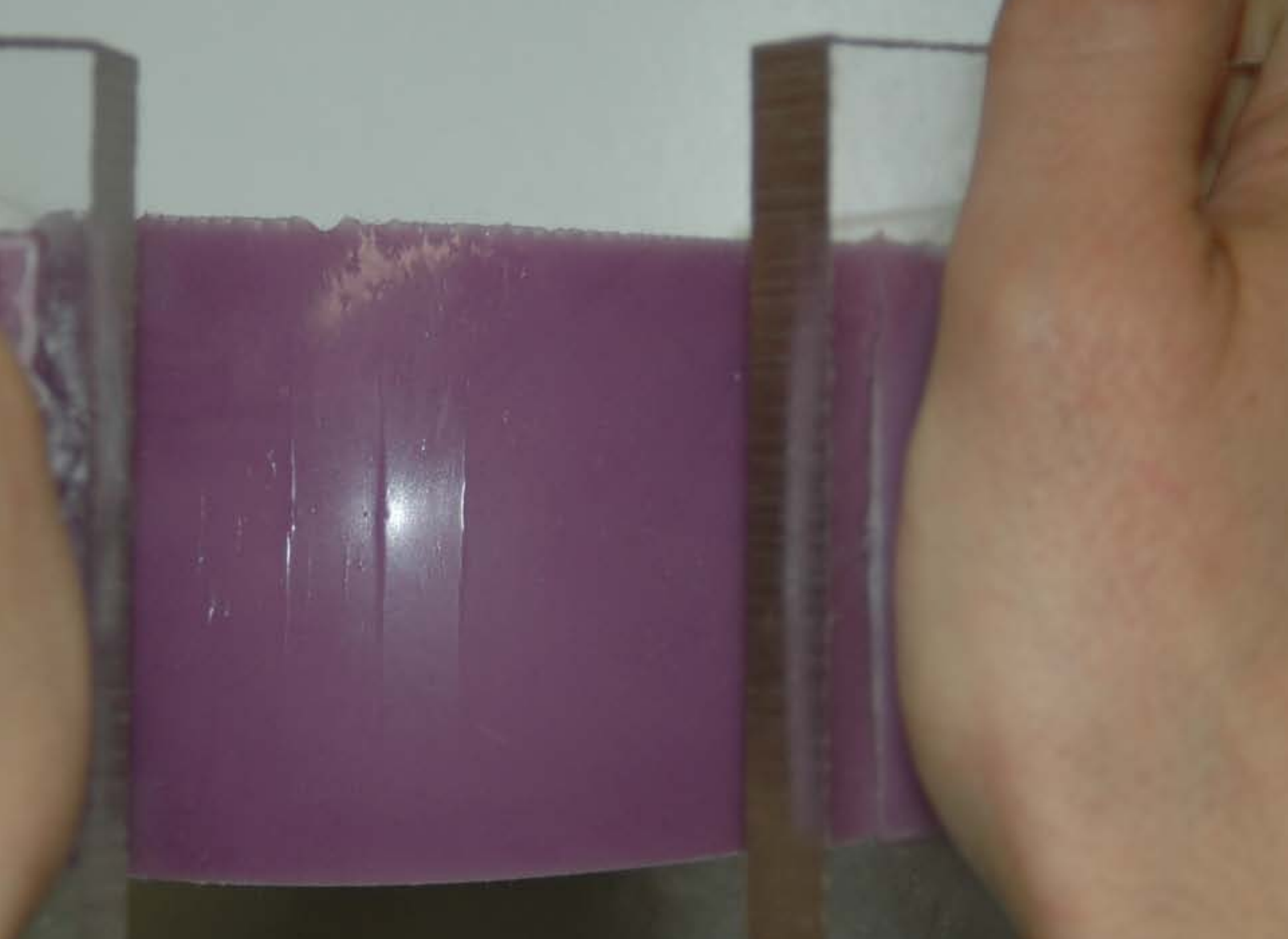}
\end{center}
\caption{Bending a block of silicone, until creases appear on the inner face. 
Here the block has dimensions $16.5$ cm $\times$ $5$ cm $\times$ $6$ cm, giving an aspect ratio $L/A=3.36$.
The end faces of the block were glued onto plexi-glass plates, and these are subjected manually to a bending moment. 
At least three axial wrinkles are visible.}
\label{fig:photo}
\end{figure}

Another deformation where the distinction between tension and compression can be determined on the basis of intuition alone is that of simple tension and compression. Subjecting a sample of rubber to a large homogeneous tension is a routine mechanical engineering experiment. Subjecting a block of rubber to a large \emph{homogeneous compression} is, on the other hand, a much more difficult task. In particular, it requires generous lubrication to avoid non-homogeneous bulging of the sides and this can lead to slippage of the block (Brown, 2005) and tilting of the platens. There is a dearth of experimental data for \emph{both} tension and compression tests of a given rubber sample. The limited data available in the literature suggest that rubbers can behave in a completely different way in one protocol compared to the other.  For example, Bechir et al. (2006) measure a Poisson ratio of 0.48 in extension for NR70 (i.e. close to incompressibility) but a Poisson ratio of 0.26 in compression. Such \emph{bimodularity} has also been observed in a variety of other elastic materials, including rocks (Lyakhovsky et al., 1997), nacre (Bertoldi et al., 2008) and cartilage (Soltz and Ateshian, 2000). A biomechanical example of bimodularity can be found in Mirnajafi et al. (2006) who measured the flexural stiffness of pig aortic valve leaflets in the direction of natural leaflet motion (Young's modulus $E^+$) and against that direction (Young's modulus $E^-$) and found that $0.43 < E^+/E^-<0.78$. Bimodularity, or tension/compression asymmetry, has long been recognized as being important in the theory of plasticity. Differences in the yield stresses between tension and compression have been observed, for example, in copper alloys (Yapici et al., 2007; Kuwabara et al., 2009) and in metallic glasses (Schuh and Lund, 2003). It is our contention that bimodularity can have a significant effect on the behavior of elastic materials as well and should be taken into account, where possible, in the analysis of their behavior.

We illustrate our argument with a study of the instability of flexure and find that it is promoted when the rubber is stiffer in compression than in tension. We take the neo-Hookean model (equivalent to Mooney-Rivlin and third-order elasticity models in bending) as the base strain-energy density. As mentioned earlier, only one material parameter plays a role in bending for these materials, namely the initial shear modulus. Therefore only $\hat \mu$, the dimensionless ratio of the shear moduli in tension and in compression, plays a role in the determination of the critical stretch of compression of bimodular third-order elasticity incompressible solids. We find that the lower and mean values of Gent and Cho's experimental range, $\lambda_\text{cr} = 0.58$ and $\lambda_\text{cr} = 0.65$, are reached for a bimodular block with $\hat \mu = 0.68$ and $\hat \mu = 0.44$, respectively. 
In general, we find that bimodularity increases the value of the critical stretch ratio of compression, and decreases the number of expected wrinkles. 
A somewhat counterintuitive outcome is that it also increases the allowable angle of bending prior to bifurcation. 
These results (\S \ref{Results}) are the product of an in-depth analysis of the large deformation field (\S \ref{Large}) and of the incremental equations of equilibrium (\S \ref{Bifurcation}).
They rely on advanced numerical methods and equations developed elsewhere (Destrade et al., 2009a).

In the concluding section (\S \ref{Conclusion}), we also present the output of a Finite Element Analysis simulation for the flexure of a uni-modular neo-Hookean block, and find that it is not satisfactory when confronted with experimental observations (Figure \ref{fig:photo}) and theoretical predictions.


\section{Bimodular bending deformation}
\label{Large}


Take a block of length $2L$, height $H$, and thickness $2A$:
\be \label{XYZ}
-A \le X \le A, \qquad -L \le Y \le L, \qquad 0 \le Z \le H,
\en
made of an incompressible, isotropic, homogeneous, non-linearly elastic \emph{bimodular} solid. Bend it under applied terminal moments into the following circular, annular sector:
\be \label{rthetaz}
r_a \le r \le r_b, \qquad -\alpha \le \theta \le \alpha, \qquad 0 \le z \le H,
\en
where  $r_a$, $r_b$ are the inner and outer radii of the bent block's curved faces, corresponding to the planes $X = -A$ and $X = A$ in the reference configuration respectively, and $2\alpha$ is the \emph{bending angle} ($0 \leq \alpha \le \pi$).
Here ($X, Y, Z$) and ($r, \theta, z$) are the rectangular and cylindrical polar coordinates of a particle before and after deformation, respectively. 
Flexure is described by Rivlin's (1949) solution
\begin{equation} \label{RivlinSln}
r = \sqrt{2 L X/\alpha +  (r_a^2 + r_b^2)/2}, \qquad 
\theta = \alpha Y / L, \qquad z = Z,
\end{equation}
from which it follows that 
\begin{equation}\label{r_ab}
r_b^2 - r_a^2 = 4 A L / \alpha.
\end{equation}

The \emph{circumferential stretch} $\lambda_2$ is a useful, non-dimensional quantity, given by
\be
\lambda_2 = \alpha r / L.
\en
If then
\be
\lambda_2^a \equiv \alpha r_a/L, \qquad 
\lambda_2^b \equiv \alpha r_b /L,
\en
we may rewrite \eqref{r_ab} as
\be \label{1st}
(\lambda_2^b)^2 - (\lambda_2^a)^2 = 2 \epsilon,
\en
where
\be
\epsilon \equiv 2 \alpha (A/ L)
\en
is a non-dimensional measure of the \emph{amount of bending}. Note that the curve separating the \emph{region of tension}, where $1 \le \lambda_2 \le \lambda_2^b$, from the \emph{region of compression}, where $\lambda_2^a \le \lambda_2 \le 1$ is called the \emph{neutral axis}, defined by, for example, Varga (1966), as
\be
r = r_n \equiv L/\alpha.
\en

We need a second relationship between $\lambda_2^a$ and $\lambda_2^b$ in order to fully determine the deformation field. The classical solution of the flexure problem due to Rivlin, where there is of course no bimodularity, assumes that the curved surfaces are stress-free and this yields $\lambda_2^a \lambda_2^b = 1$, which is independent of the strain-energy function. In a \emph{bimodular} block, the stress-free boundary condition yields a second relationship that does depend on the form of the strain-energy function assumed. Here we assume that
\be \label{sef}
W^\pm = \mu^\pm\left(\lambda_2^{-2}+\lambda_2^2-2\right),
\en
where $\mu^\pm$ are the shear moduli in tension and compression respectively. This is the plane strain form of the neo-Hookean strain-energy function and we recall from the Introduction that a wide class of materials reduces to this form in plane strain. For this material, Destrade et al. (2009b) show that when the curved faces of the bent block are free of normal traction and the stress is continuous throughout the bimodular block, then the following equation is obtained:
\begin{equation} \label{2nd}
2\epsilon  x^3+\left[1-4\epsilon-\hat{\mu}\left(1-2\epsilon\right)^2\right]x^2 - 2\left[1 - \epsilon - \hat{\mu}\left(1-2\epsilon\right)\right]x + 1-\hat{\mu}=0,
\end{equation}
where $x\equiv(\lambda_2^a)^{-2}$  and $\hat{\mu}\equiv \mu^+ /\mu^-$ is the ratio of the shear moduli in the regions of tension and compression, respectively. 
This is a bicubic in $\lambda_2^a$ and a quadratic in $\epsilon$.

\begin{figure}
\begin{center}
\includegraphics*[width=10cm]{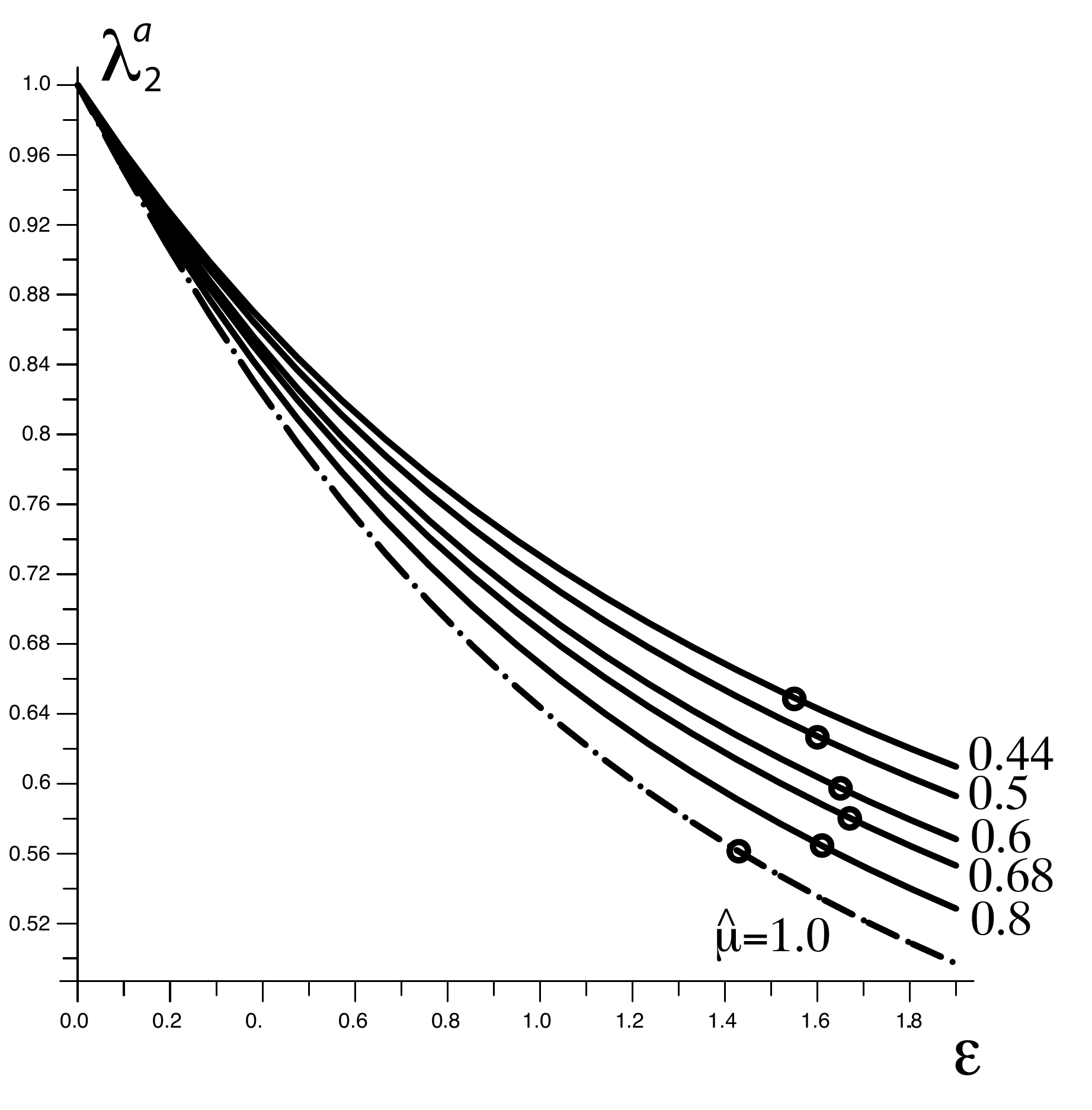}
\end{center}
\caption{Variations of $\lambda_2^a$, the circumferential stretch ratio on the inner bent face,
with $\epsilon$, the product of the block aspect ratio and the bending angle, for bimodular solids which are stiffer in compression than in tension. 
The ratio of the tension shear modulus to the compression shear modulus takes the values: 
$\hat \mu = 1.0$ (dotted line; homogeneous block), $0.8$, $0.68$, $0.6$, $0.5$, and $0.44$ (full lines).
The circles give an indication of the critical stretch of bending instability, see Figure \ref{fig:main}.}
\label{fig:lambda_2}
\end{figure}

Equations \eqref{1st} and \eqref{2nd} determine completely the large bending deformation for a bimodular neo-Hookean block,  with a given $\hat \mu$, once the amount of bending $\epsilon = 2\alpha (A/L)$ is prescribed, since $\lambda_2^a$ can be determined from \eqref{2nd} and $\lambda_2^b$ then obtained from \eqref{1st}. To illustrate the magnitude of the stretches obtained, Figure \ref{fig:lambda_2} shows the variations of $\lambda_2^a$ with $\epsilon$. 

Once $\epsilon$, $\lambda_2^a$, and $\lambda_2^b$ are known, the inner and outer radii, as well as the neutral axis, are completely determined. 
Normalized with respect to the block thickness, they are
\be
\dfrac{r_a}{2A} = \dfrac{\lambda_2^a}{\epsilon}, \qquad
\dfrac{r_n}{2A} = \dfrac{1}{\epsilon}, \qquad
\dfrac{r_b}{2A} = \dfrac{\lambda_2^b}{\epsilon}.
\en

Finally, the radial stress field is either $\mu^+ \sigma^+$ or $\mu^- \sigma^-$, depending on the region---tension or compression.
Here, the non-dimensional quantities $\sigma^\pm$ are 
\be
\sigma^+ = \halft\left[\lambda_2 + \lambda_2^{-2} - (\lambda_2^b)^2 - (\lambda_2^b)^{-2}\right],
\qquad
\sigma^- = \halft\left[\lambda_2 + \lambda_2^{-2} - (\lambda_2^a)^2 - (\lambda_2^a)^{-2}\right].
\en
This field leaves the bent faces free of normal traction.


\section{Bimodular bending instability}



\subsection{The general method}


Suppose that instead of specifying $\epsilon$, as in the last section, $\lambda_2^a$, the circumferential stretch on the inner face, is prescribed. Then \eqref{2nd} can be solved to determine $\epsilon$, with  $\lambda_2^b$ determined from \eqref{1st}. We can then  investigate whether, at the prescribed stretch $\lambda_2^a$, a static incremental field can be superimposed onto the large bending solution. If so, then $\lambda_2^a$ is what is termed a \emph{critical stretch of bending instability}, $\lambda_\text{cr}$. The incremental equations of equilibrium used to determine these critical stretches are described next.


\subsection{Bifurcation}
\label{Bifurcation}


For incompressible solids, the incremental equations of equilibrium are (Ogden, 1984)
\be \la{equi}
\text{div } \vec{\dot{s}} = \mathbf{0}, \qquad 
\text{div } \vec{u} = 0,
\en
where $\vec{\dot{s}}$ is the incremental nominal stress
and $\vec{u}$ is the incremental mechanical displacement.
We seek solutions of these equations in the form
\be \la{soln}
\{\vec{u}, \vec{\dot{s}} \} = \Re\left\{\left[U(\lambda_2), V(\lambda_2), 0, S_{rr}(\lambda_2), S_{r\theta}(\lambda_2), 0\right] \ee^{\ii n\theta}\right\}, 
\en
where $U$, $V$,  $S_{rr}$, $ S_{r\theta}$, are complex functions of $\lambda_2$ only and
\be \la{ndef}
n = p \pi / \alpha, \qquad \text{ $p$ an integer},
\en
gives the number of wrinkles on the bent faces.

Destrade et al. (2009a) formulated the following Stroh form of the incremental equations:
\be \la{stroh}
\dfrac{\text{d}}{\text{d} \lambda_2} \vec{\eta}(\lambda_2) = \dfrac{\ii}{\lambda_2} \vec{G^\pm}(\lambda_2) \vec{\eta}(\lambda_2),
\en
where
\be
\vec{\eta} \equiv [U, V, \ii r S_{rr}, \ii r S_{r \theta}]^t
\en
is the displacement-traction vector (Shuvalov, 2003). The components of the matrices $\vec{G^\pm}$ are given by
\be \label{G}
\vec{G^\pm} = \begin{bmatrix} 
  \ii & -n & 0 & 0 \\
  -n (1 - \sigma^\pm \lambda_2^2) &  - \ii (1 - \sigma^\pm \lambda_2^2) & 0  &  - \lambda_2^2/\mu^\pm \\
  \kappa_{11}^\pm & \ii \kappa_{12}^\pm & -\ii & -n (1 - \sigma^\pm \lambda_2^2) \\
  - \ii \kappa_{12}^\pm & \kappa_{22}^\pm & -n &  \ii (1 - \sigma^\pm \lambda_2^2)  
          \end{bmatrix}
\en
where
\begin{align}
& \kappa_{11}^\pm = \mu^\pm \left\{\lambda_2^2 + 3\lambda_2^{-2} - 2\sigma^\pm  
  + n^2\lambda_2^2\left[1 - (\lambda_2^{-2} - \sigma^\pm)^2\right] \right\},
 \notag \\[2pt]
& \kappa_{12}^\pm = n \mu^\pm\left[2 \lambda_2^2 + 2\lambda_2^{-2} - \lambda_2^2(\sigma^\pm)^2\right],
 \notag \\[2pt]
& \kappa_{22}^\pm =  
\mu^\pm   \left[ \lambda_2^2 - \lambda_2^2(\lambda_2^{-2} - \sigma^\pm)^2
   + n^2(\lambda_2^2 + 3\lambda_2^{-2} - 2\sigma^\pm) \right].
\end{align}

The bifurcation problem is a two-point boundary value problem. Thus \eqref{stroh} must be integrated numerically, subject to the following boundary conditions of zero incremental traction on the bent faces:
 \be \la{BC}
 \vec{\eta} (\lambda_2^a) = [\vec{U}(\lambda_2^a), \vec{0}]^t,
\qquad 
 \vec{\eta} (\lambda_2^b) = [\vec{U}(\lambda_2^b), \vec{0}]^t.
\en
Following the approach of Haughton (1999), the value of $n$, defined in \eqref{ndef}, is determined by assuming that there are no incremental normal tractions on the plane end-faces $\theta = \pm \alpha$. We must also enforce the continuity of displacement and traction (i.e. of $\vec{\eta}$) across the neutral axis (at $\lambda_2=1$). Several strategies exist to tackle this problem.
To find $\lambda_\text{cr}$ only, the \emph{compound matrix method} is most efficient;
to find $\lambda_\text{cr}$ \emph{and} the mechanical fields throughout the block, the \emph{impedance matrix method} works very well.
In the former case, we have to integrate numerically a linear system satisfied by the compound matrix; in the latter case, we have to integrate numerically a Riccati equation satisfied by the conditional surface impedance matrix (see Destrade et al. (2009a) for details).


\subsection{Results}
\label{Results}


For a given bimodular material, $\hat \mu = \mu^+/\mu^-$ is prescribed. Non-di\-men\-sion\-a\-li\-sing the incremental equations \eqref{stroh} yields two independent, non-dimensional parameters: $\lambda_\text{cr}$ and $L/(pA)$. Plotting then $\lambda_\text{cr}$ versus $L/(pA)$ yields master curves for the ``physical'' dispersion curves, which give the critical stretch ratio $\lambda_\text{cr}$ versus the block aspect ratio $L/A$, for $p=1,2,3,\ldots$. The maximum of the master curve is a good indicator of the physical critical stretch of a block with a given aspect ratio. Further details can be found in Haughton (1999), Coman and Destrade (2008), and Destrade et al. (2009a).

\begin{figure}
\begin{center}
\includegraphics*[width=10cm]{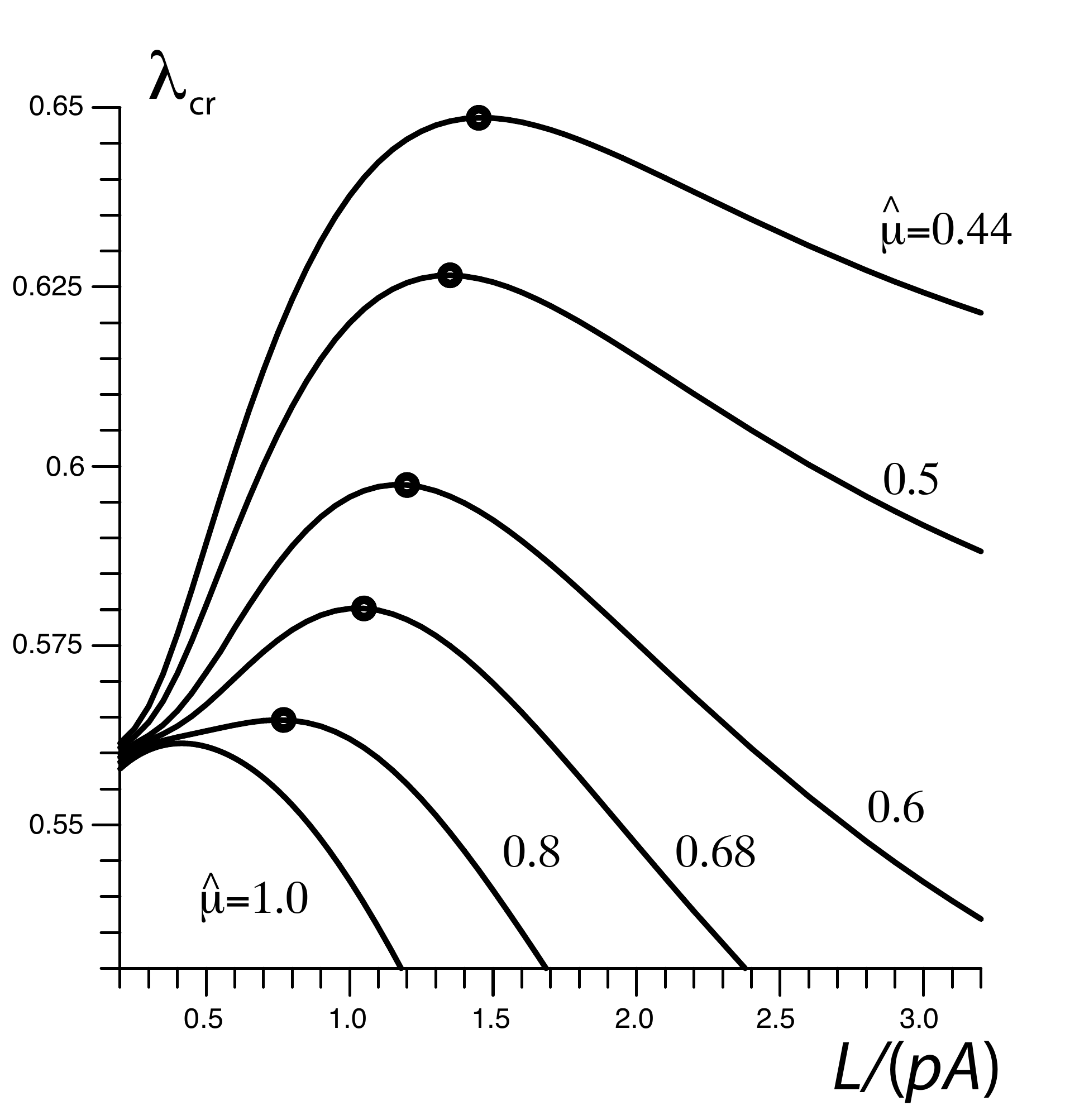}
\end{center}
\caption{How bimodularity promotes bending instability. As the difference between compressive and tensile behaviors becomes more pronounced ($\hat \mu = 1.0, 0.8, \ldots, 0.44$), the critical stretch of bending increases. (We checked separately that when the solid is stiffer in tension than in compression, the block becomes more stable.) The circles indicate the maximum of each curve, and give a good indication of the ``physical critical stretch'' of bifurcation, see Figure \ref{fig:example}.}
\label{fig:main}
\end{figure}

Consequently, we plot these master curves for various values of $\hat \mu$ within the range $0.44 \le \hat \mu \le 1$; the upper value restricts attention to bimodular solids which are stiffer in compression than in tension, the choice of the arbitrary lower value will be explained shortly. Qualitatively, we find that the bending instability occurs earlier for bimodular solids than for a uni-modular block ($\hat \mu = 1$), in line with results on coated blocks (Dryburgh and Ogden, 1999) and bilayered blocks (Roccabianca and Bigone, private communication). Quantitatively, we find that the lower value of the Gent and Cho (1999) range, i.e. $\lambda_\text{cr} = 0.65 - 0.07 = 0.58$, corresponds to $\hat \mu = 0.68$ whilst their mean value, $\lambda_\text{cr} = 0.65$ corresponds to $\hat \mu = 0.44$, see Figure \ref{fig:main}.
On that figure, we note that the maximum of each master curve increases as the bimodularity becomes more pronounced.

There is a seemingly counter-intuitive result worth noting: although instability is clearly promoted by bimodularity, in the sense that $\lambda_\text{cr}$ is increased, a bimodular block can nonetheless be bent further than its uni-modular counterpart before instability occurs, given that instability for a bimodular block corresponds to a larger value of the amount of bending $\epsilon = 2 \alpha (A/L)$. This is clear from the observation of the circles on Figure \ref{fig:lambda_2}, which indicate the onset of bending instability. This occurs not alone at a higher value of $\lambda_2$ than the homogeneous (uni-modular) block (at $\hat \mu = 1$), but also at a higher value of $\epsilon$. We illustrate this observation further with a specific example.

\begin{figure}
\begin{center}
\includegraphics*[width=11cm]{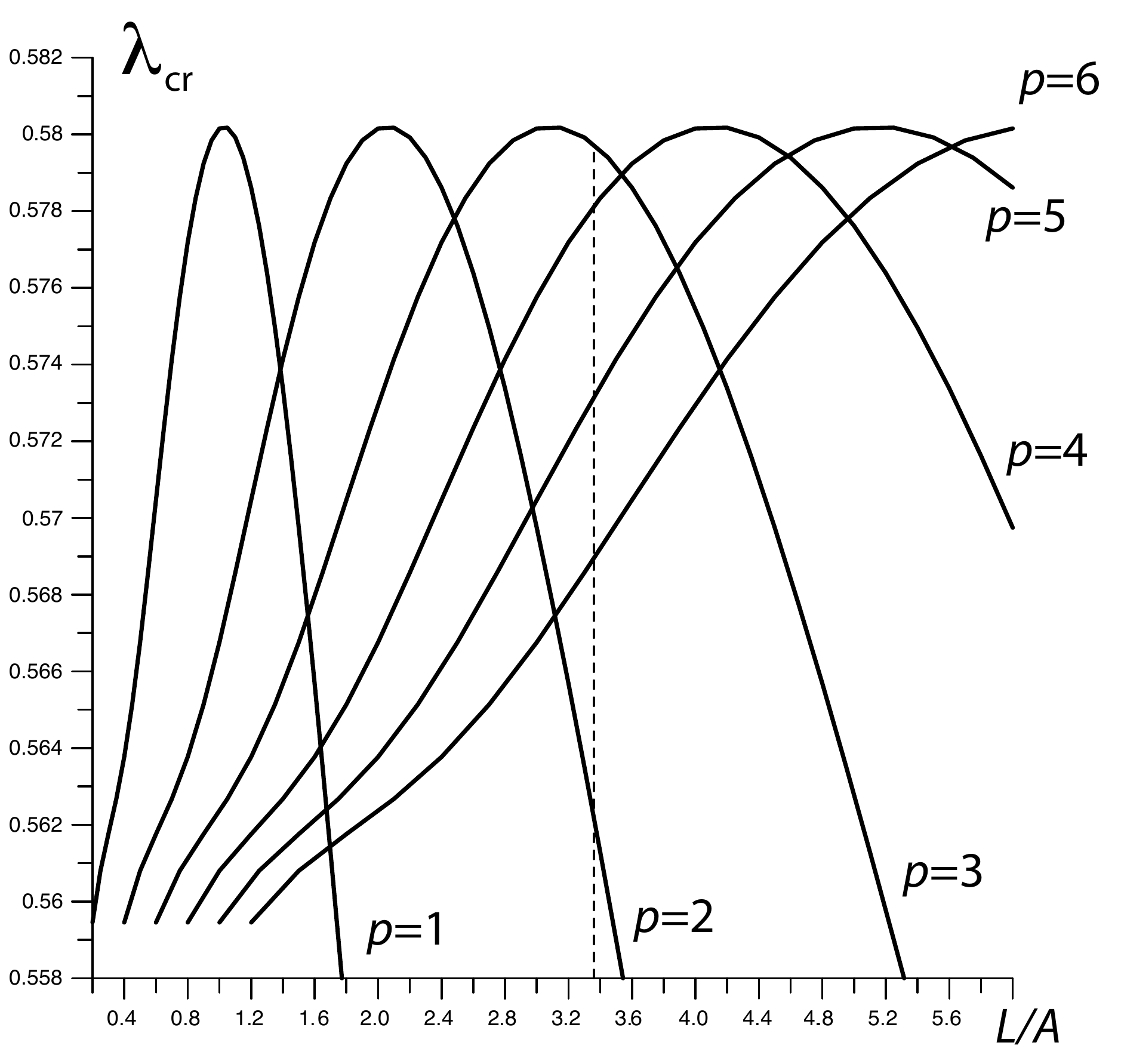}
\end{center}
\caption{Dispersion curves for a given bimodular block, with $\hat \mu = 0.68$.
Depending on the aspect ratio $L/A$ of the block, a certain mode of buckling is triggered at $\lambda_\text{cr}$.
Except for unrealistically short and stubby blocks, $\lambda_\text{cr}$ is confined within the range $0.578-0.580$, that is, close to the value indicated by the circle on Figure \ref{fig:main}.
For example, a block with aspect ratio $L/A=3.36$ (like the silicone block of Figure \ref{fig:photo}) experiences bending instability when the circumferential stretch reaches 0.5797 approx. and three axial wrinkles should appear ($p=3$), as can be easily seen from consideration of the vertical hash line.}
\label{fig:example}
\end{figure}

Figure \ref{fig:example} displays the dispersion curves $p=1,2, \ldots, 6$ when $\hat \mu = 0.68$.
When the aspect ratio $L/A = 3.36$ (this is the aspect ratio of the silicone block used in Figure \ref{fig:photo}), it is clear from the figure that its inner face buckles in bending when $p=3$ and $\lambda_2^a = 0.5797$.

\begin{figure}
\begin{center}
\includegraphics*[width=14cm]{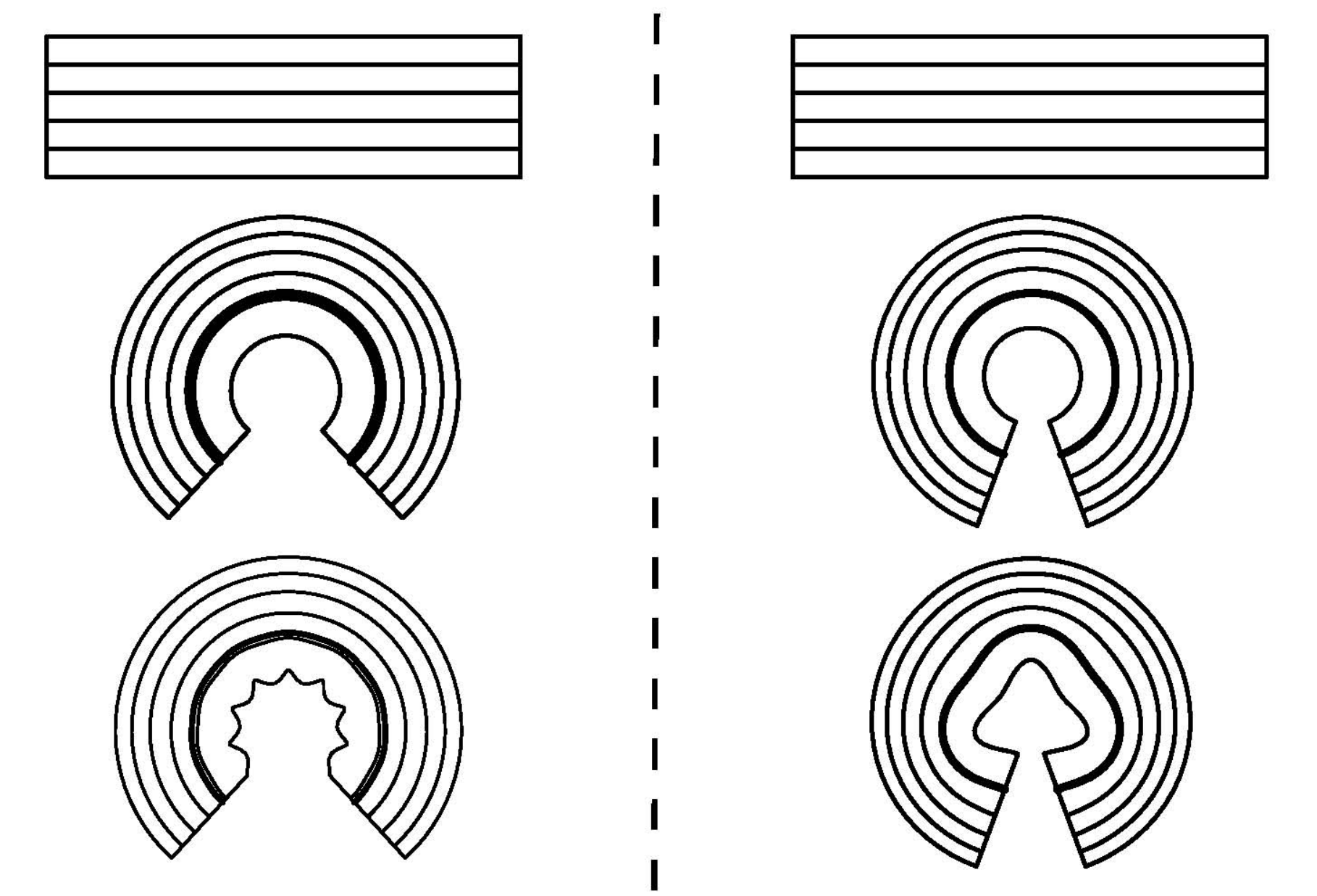}
\end{center}
\caption{Mechanical displacement field in bent blocks with aspect ratio $L/A=3.36$ (like the block of Figure \ref{fig:photo}). On the left: uni-modular block ($\hat \mu = 1$); on the right: a bimodular block with $\hat \mu = 0.68$.
First stage: undeformed configuration.
Second stage: maximum bending, prior to bifurcation; here the angles of bending are $275^\circ$ and $319^\circ$, for the uni-modular and the bimodular blocks, respectively, corresponding to critical stretches of $0.56$ and $0.58$, respectively.
Third stage: incremental buckling; here 8 wrinkles and 3 wrinkles appear on the inner face of the uni-modular block and of the bimodular block, respectively.
The thick line is the neutral axis, where line elements are neither extended nor contracted.}
\label{fig:three_stages}
\end{figure}

Figure \ref{fig:three_stages} contrasts this incremental buckling field throughout the bent block (right-hand side of the figure) with the corresponding uni-modular field. From \eqref{2nd}, the corresponding amount of bending for the bimodular block is $\epsilon = 1.660$, giving a bending angle of $2\alpha = \epsilon(L/A) = 319^\circ$.
In comparison, a uni-modular block ($\hat \mu = 1$) with the same dimensions buckles in bending with $p=8$ at $\lambda_\text{cr} = 0.5612$ (left side of the figure). 
Hence, by moving the maximum of the master curve upwards and to the right on Figure \ref{fig:main}, bimodularity promotes earlier buckling ($\lambda_\text{cr}$ increases), with fewer wrinkles ($p$ decreases). 
At the same time, $\epsilon$ increases too (see Figure \ref{fig:lambda_2}), which means that the critical bending angle is larger.
Note that Figure \ref{fig:three_stages} displays the correct (scaled) size of the bent blocks relative to the undeformed blocks. 
Similarly, the amplitude of the incremental wrinkles on the outer face is computed relative to their amplitude on the inner face. 
However, the amplitude of the wrinkles relative to the large bending displacement is not known, because the incremental analysis is linear.

\section{Concluding remarks}
\label{Conclusion}


Gent and Cho (1999) showed experimentally that a block of rubber buckles in bending when the circumferential stretch is $0.65 \pm 0.07$, earlier than their prediction based on surface buckling, where the critical stretch is 0.54. 
Previous modeling attempts at finding a satisfactory explanation for this early buckling have failed. 
They included taking into account slight compressibility, finite dimensions, and the strain-stiffening effect. 
Here we investigated how \emph{bimodularity} might affect the bending instability of rubber blocks, and found that if the rubber is stiffer in compression than in tension, then the bending buckling does occur earlier. 
Our detailed analysis also revealed  that the bimodular block can be bent by a larger angle than the corresponding uni-modular block with the same dimensions. 
It was important to evaluate the effect of bimodularity for the flexure stability problem, because numerical simulations using the Finite Element Method do not provide physical and realistic predictions. We conclude the paper by summarizing our numerical simulation experiments.

\begin{figure}
\centering
 \mbox{\subfigure{\epsfig{figure=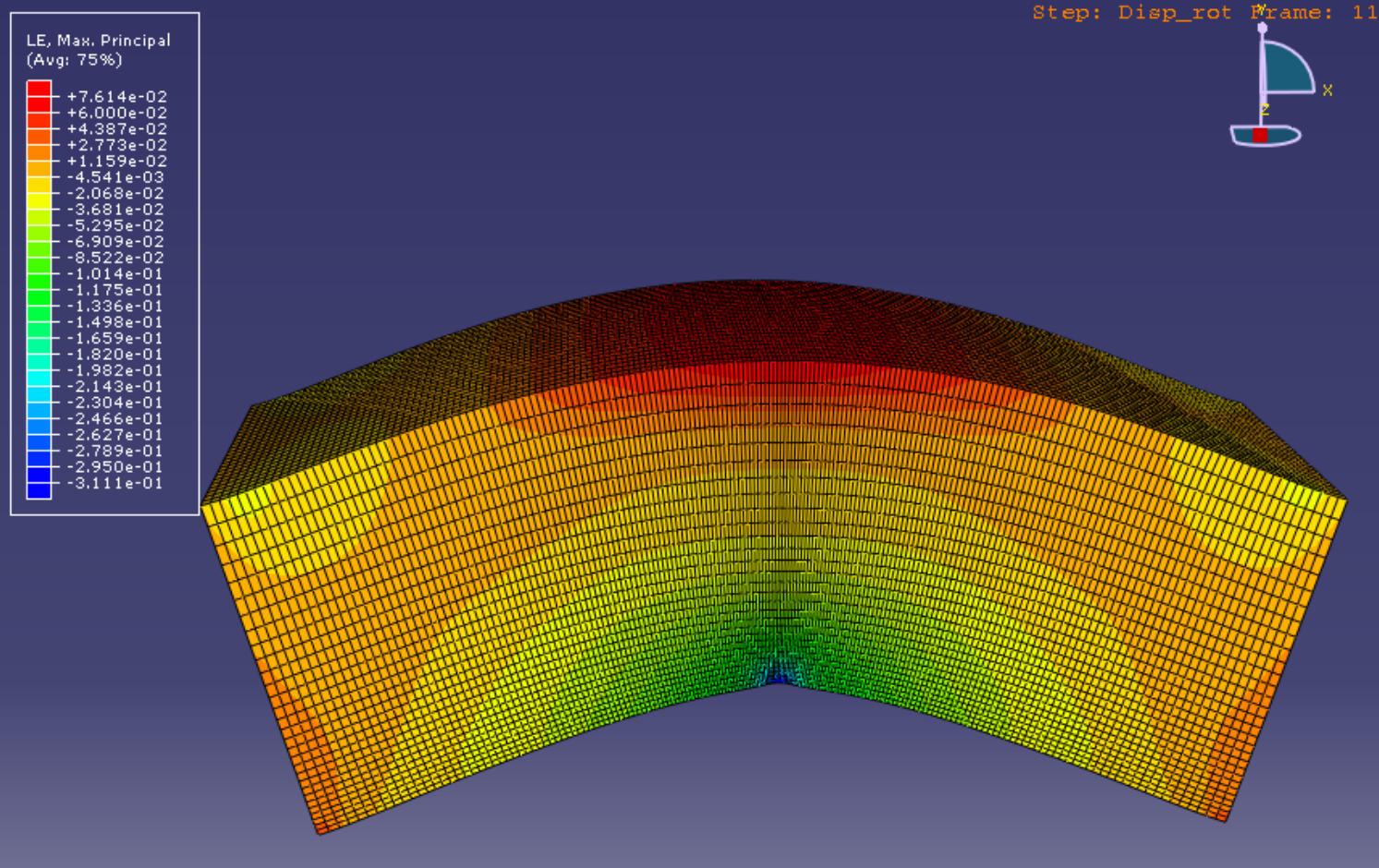, width=.45\textwidth}}}
  \quad \quad
     \subfigure{\epsfig{figure=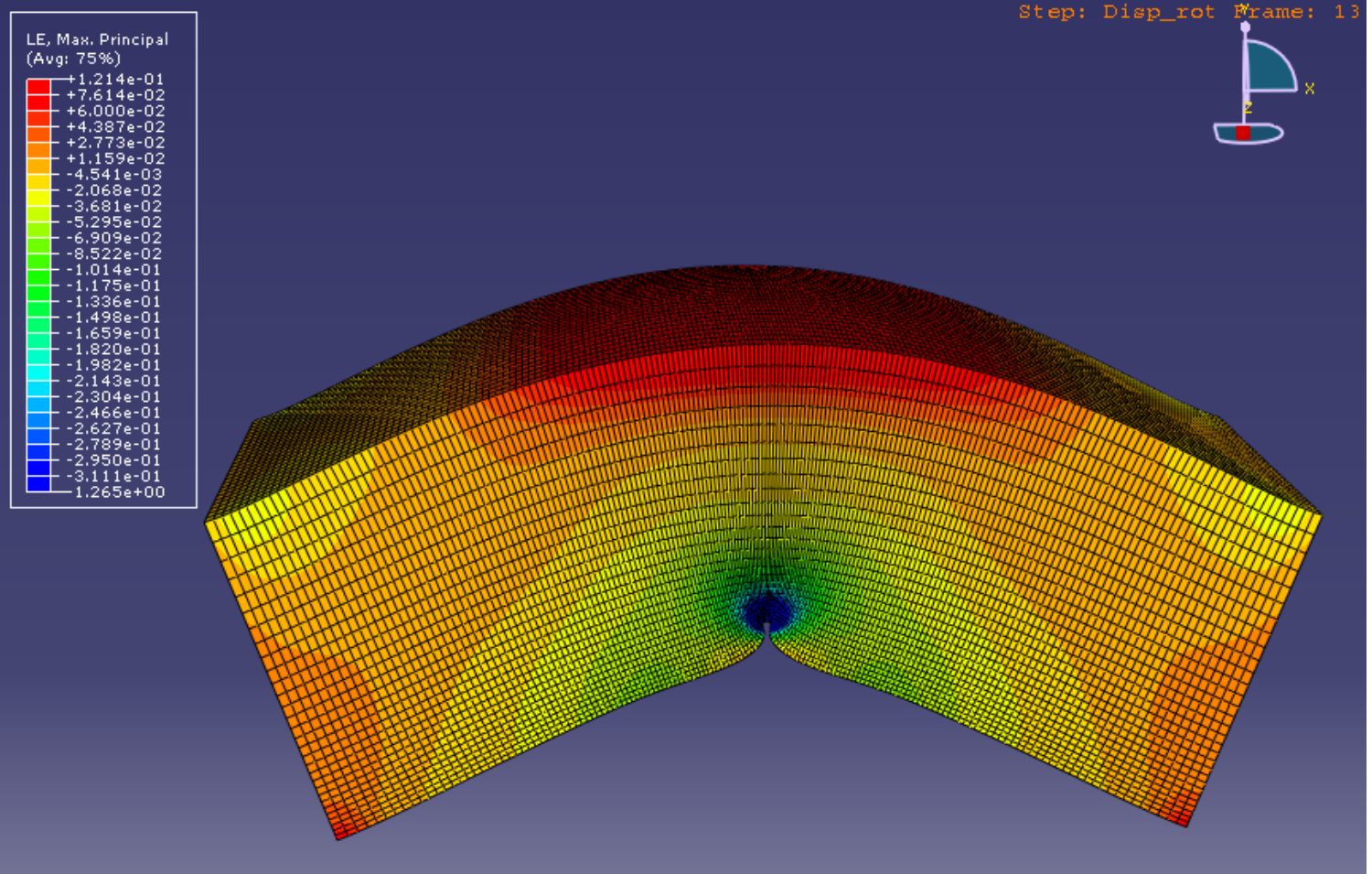, width=.45\textwidth}}
\\
 \mbox{\subfigure{\epsfig{figure=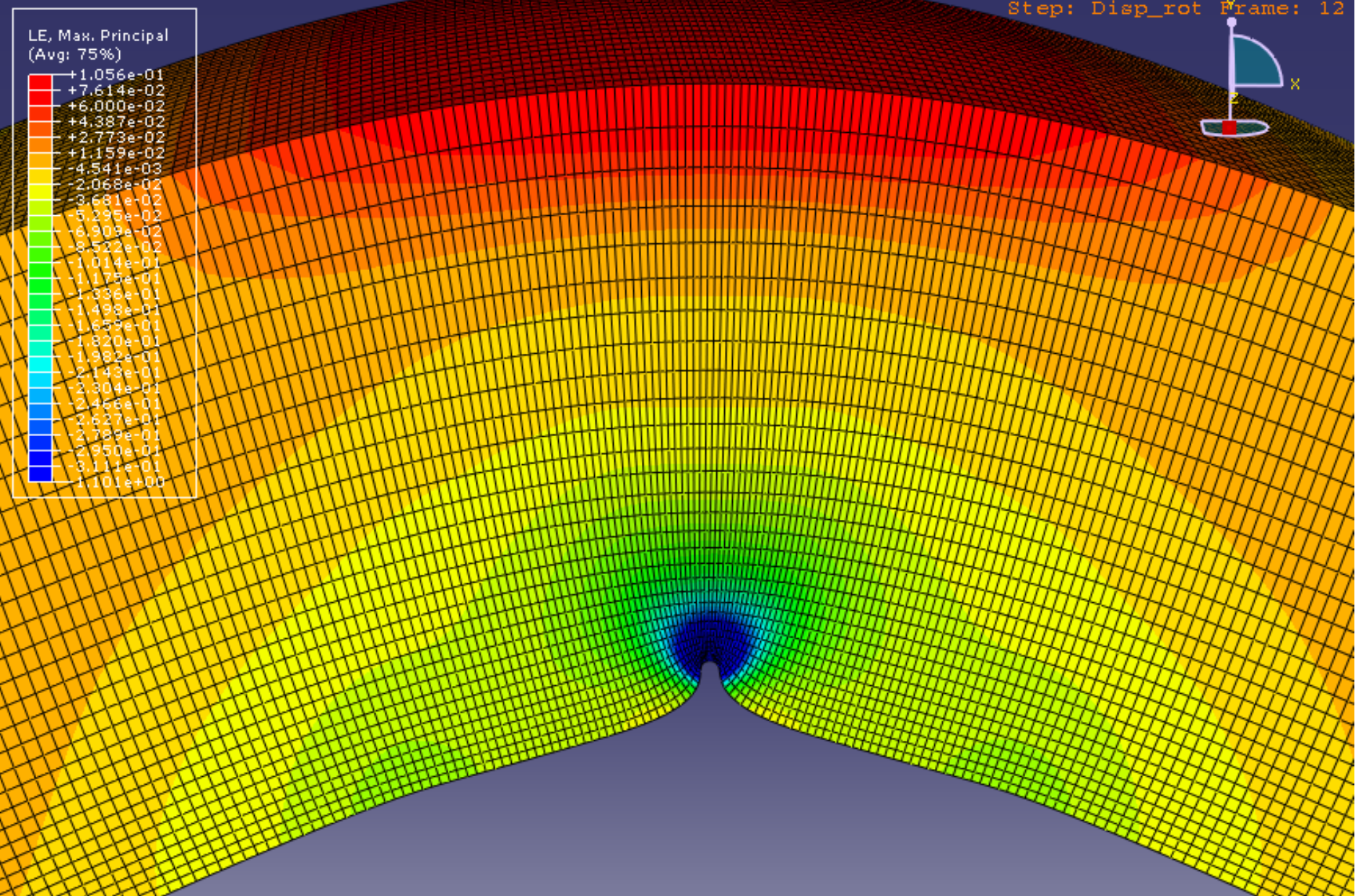, width=.45\textwidth}}}
  \quad \quad
     \subfigure{\epsfig{figure=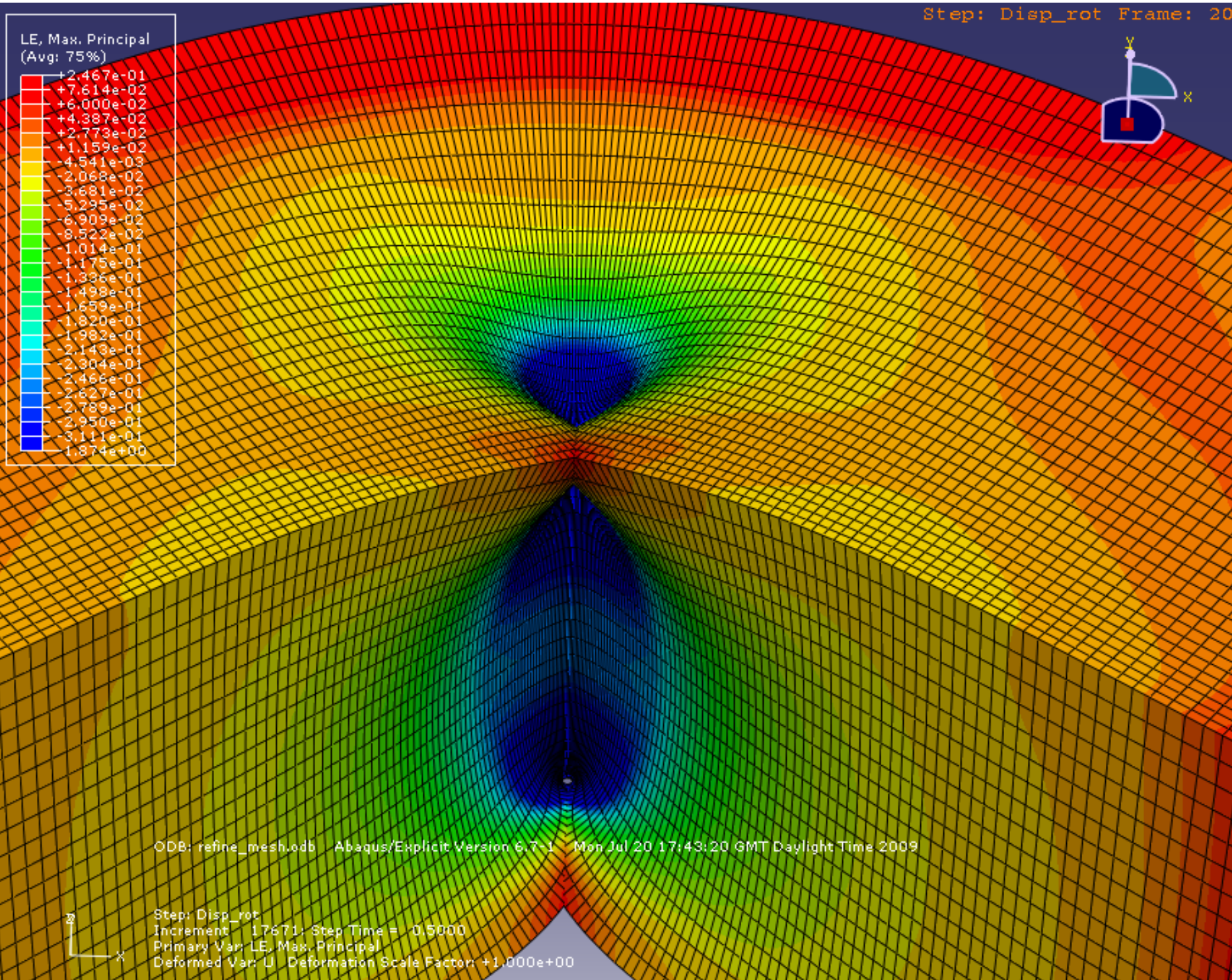, width=.45\textwidth}}
 \caption{Finite element simulation of the large flexure of a block with the same dimensions as the block of Figure \ref{fig:photo}.
 The top pictures shows the results of the simulation just prior to folding of the inner surface (first picture) and after (second picture). 
 The bottom pictures show zooms on the fold in the buckled geometry.}
 \label{fig:FEM}
\end{figure}

We used the commercial Finite Element Analysis software Abaqus [Dassault Syst\`emes Simulia Corp., Providence, Rhode Island, USA] to simulate numerically the flexure of a uni-modular block with the same dimensions (16.5 cm $\times$ 5 cm $\times$ 6 cm) as the silicone block of Figure 1. We measured its mass density as being $\rho = 641.176$ kg.m$^{-3}$.
A hyperelastic material model was chosen using the neo-Hookean form of the strain energy potential defined as follows: initial shear modulus $\mu =  0.6$ MPa and initial Poisson ratio 0.4999 (to mimic incompressibility). 
A reference point at the center of each of the respective end faces of the block governed the movement of each of these faces by means of a coupling constraint. 
Each of the reference points was displaced 3.91 cm towards the other, and they were rotated by $60^\circ$. 
These values were computed as being consistent with Rivlin's plane strain solution \eqref{RivlinSln}. 
The displacement and rotation were ramped linearly over a time step of 0.5 s. 
The mesh consisted of first-order fully integrated 8 noded hexahedral 3D continuum elements, enhanced by incompatible modes to improve their bending behavior. The analysis was computed using explicit procedures to allow for the large deformation.
We looked at different types of element formulations and mesh refinements, but always found the same following trends in the final output. 
As shown on Figure \ref{fig:FEM}, Abaqus predicts that under severe bending, the block should experience the formation of a single fold, centered axially on the inner bent face. This fold occurs at a smaller bending angle, and at larger strains, than observed experimentally and predicted theoretically. Other inconsistencies include the absence of other folds, the opposite of the anti-clastic effect (in the sense that the outer face widens while the inner face shrinks), and the penetration of the fold inside the block (as seen on the last picture).

\section*{Acknowledgments}


This work is supported 
by a Marie Curie Fellowship for Career Development awarded by the Seventh Framework Programme (European Commission) to the first author.



\end{document}